\documentclass[12pt]{article}
\usepackage{amssymb}
\usepackage{amsmath}
\usepackage{epsfig}
\usepackage{float}
\newcommand{\be}{\begin{equation}}
\newcommand{\ee}{\end{equation}}
\newcommand{\bea}{\begin{eqnarray}}
\newcommand{\eea}{\end{eqnarray}}
\begin{document}

\begin{center}
{\bf ARE NEUTRINO OSCILLATIONS A NON-STATIONARY
PHENOMENON?}\footnote{To be published in the Proceedings of the XIII. International workshop on "Neutrino Telescopes", Venice, March 10-13, 2009; edited by Milla Baldo Ceolin.}
\end{center}

\begin{center}
S. M. Bilenky
\end{center}

\begin{center}
{\em  Joint Institute for Nuclear Research, Dubna, R-141980,
Russia\\}
{\em Physik-Department E15, Technische Universit\"at M\"unchen,
D-85748 Garching, Germany}
\end{center}
\begin{center}
F. von  Feilitzsch and W. Potzel
\end{center}
\begin{center}
{\em Physik-Department E15, Technische Universit\"at M\"unchen,
D-85748 Garching, Germany}
\end{center}

\begin{abstract}
We discuss different schemes of neutrino
oscillations and the time-energy uncertainty relation.  From
the results of the K2K and MINOS accelerator experiments
follows that neutrino oscillations are a non-stationary
phenomenon and that the standard time-energy uncertainty
relation is satisfied. Are these properties a general feature
of neutrino oscillations? In this paper we demonstrate that a
recently proposed tritium/helium-3 M\"ossbauer
neutrino-experiment could answer this question.
\end{abstract}

\section{Introduction}
 The observation of neutrino oscillations in the Super
 Kamiokande at\-mo\-spheric \cite{SK}, SNO solar \cite{SNO},
 KamLAND reactor \cite{Kamland}, and other neutrino
 experiments \cite{other,K2K,Minos} is one of the most
 important recent discoveries in particle physics. It is
 common opinion that small neutrino masses and peculiar
 neutrino mixing are signatures of physics beyond the
 Standard Model. The principal aim of future experiments
 concerning the study of neutrino oscillations, the search
 for neutrinoless double $\beta$-decay and the direct
 measurement of neutrino masses is to make a further step in
 the investigation of neutrino properties which would allow
 to reveal new physics determining the nature of neutrinos
 and of small neutrino masses.

 In the
T2K \cite{T2K}, Double CHOOZ \cite{2chooz}, Daya Bay
\cite{Dayabay} and other next-gen\-er\-ation neutrino oscillation
experiments, a high accuracy in the measurement of
oscillation parameters will be achieved. An even higher
accuracy is planned to be reached in experiments at future
Super beam, $\beta$-beam and Neutrino Factory facilities (see
Ref. \cite{NF}).

We will address the following related questions:
\begin{itemize}
  \item Are neutrino
oscillations a non-stationary phenomenon?
  \item Are stationary neutrino oscillations also
      possible?
\end{itemize}
\section{Neutrino mixing}

The investigation of neutrino oscillations is based on the
following assumptions (see, for example, Refs.
\cite{BilPont,BilPet,BGG}):
\begin{enumerate}
  \item The neutrino processes are described by the SM
      charged current (CC) and neutral current (NC)
      interactions. The Lagrangian of the standard
      leptonic CC interaction is given by the expression
\begin{equation}\label{CCL}
\mathcal{L_{I}}^{CC}(x)=-\frac{g}{2\sqrt{2}}~
j_{\alpha}^{CC}(x)~W_{\alpha}(x)+\mathrm{h.c.}~,
\end{equation}
 where
\begin{equation}\label{CCL1}
j_{\alpha}^{CC}(x)=2\, \sum_{l=e,\mu,\tau}\bar \nu_{l
L}(x) \,\gamma
_{\alpha}\, l_{L}(x)~.
\end{equation}
\item The field $\nu_{l L}(x)$ is a unitary "mixture" of
    left-handed components of fields of Majorana (or
    Dirac) neutrinos $\nu_{i}$  with mass $m_{i}$:
\begin{equation}\label{CCL2}
\nu_{l L}(x)=\sum^{3}_{i=1} U_{l i}\,\nu_{i L}(x)~.
\end{equation}
Here $U$ is the unitary PMNS mixing matrix.
\end{enumerate}
The relation (\ref{CCL2}) is a relation between  quantum
fields. There is no dispute  about this relation. There have
been, however, many discussions on the consequences which
follow from this relation for the observable neutrino
transition probabilities (see Ref. \cite{Giunti} and
references therein).

\section{Neutrino oscillation data}

All existing neutrino oscillation data can be described if we
assume that
\begin{enumerate}
  \item The number of massive neutrinos is equal to the
      number of flavor neutrinos (three).
  \item The  $\nu_{l}\to\nu_{l'}$  transition probability
      is given by the expression
\begin{equation}\label{transprob}
P(\nu_{l}\to\nu_{l'})=|\sum^{3}_{i=1}U_{l'i}~ e^{-i\Delta
m_{ki}^{2}\frac{L}{2E}}~ U^{*}_{li}|^{2}~.
\end{equation}
Here $\Delta m_{ki}^{2}= m_{i}^{2}- m_{k}^{2}$, $L$ is
the distance between the neutrino source and the neutrino
detector, $E$ is the neutrino energy, and the index $k$
is fixed.
\end{enumerate}
The probabilities $P(\nu_{l}\to\nu_{l'})$ depend on six
parameters (two mass-squared differences $\Delta m_{12}^{2}$
and
$\Delta m_{23}^{2}$, three mixing angles $\theta_{12},
\theta_{23}$ and $\theta_{13}$, one phase  $\delta$) and have
rather complex forms.
However, we can take into account that two parameters are
small:
\begin{equation}\label{smallpar}
\frac{\Delta m^{2}_{12}}{\Delta m^{2}_{23}}\simeq 3\cdot
10^{-2},\quad \sin^{2}\theta_{13}\lesssim 5\cdot 10^{-2}~.
\end{equation}
If we neglect the contributions of these small parameters to
the transition probabilities and consider the range of
$\frac{L}{E}$-values characteristic for LBL
atmospheric-neutrino experiments ($\Delta
m_{23}^{2}\frac{L}{2E}\gtrsim 1$), the leading oscillations
are, in this case,
$\nu_{\mu}\leftrightarrows \nu_{\tau}$ (see, for example,
Ref. \cite{BGG}). For the probability of
$\nu_{\mu}$ ($\bar\nu_{\mu}$) to survive we find the
following expression
\begin{equation}\label{transprob1}
P(\nu_{\mu}\to\nu_{\mu})=P(\bar\nu_{\mu}\to\bar\nu_{\mu})\simeq
1-
\frac{1}{2}\sin^{2}2\theta_{23}
(1-\cos\Delta m_{23}^{2}\frac{L}{2E})~.
\end{equation}
For the values of  $\frac{L}{E}$ which satisfy the condition
$\Delta m_{12}^{2}\frac{L}{2E}\gtrsim 1$ (KamLAND range),
in leading approximation  the probability of $\bar\nu_{e}$ to
survive takes the two-neutrino form
\begin{equation}\label{transprob2}
P(\bar\nu_{e}\to \bar\nu_{e})\simeq
1-\frac{1}{2}\sin^{2}2\theta_{12}
(1-\cos\Delta m_{12}^{2}\frac{L}{2E})~.
\end{equation}
Notice that both  $\bar\nu_{e}\to\bar\nu_{\mu}$ and
$\bar\nu_{e}\to\bar\nu_{\tau}$ transitions contribute to
$P(\bar\nu_{e}\to \bar\nu_{e})$.

In leading approximation, also the probability of the solar
$\nu_{e}$'s to survive in  matter has the two-neutrino form
and depends on the parameters $\Delta m_{12}^{2}$ and
$\sin^{2}\theta_{12}$.

From the
analysis of the data of the S-K atmospheric neutrino
experiment, the following 90\% CL ranges were obtained
\cite{SK} for the parameters $\Delta m^{2}_{23}$
and $\sin^{2}2 \theta_{23}$
\begin{equation}\label{SKrange}
 1.9\cdot 10^{-3}\leq \Delta m^{2}_{23} \leq 3.1\cdot
10^{-3}\rm{eV}^{2},\quad \sin^{2}2 \theta_{23}> 0.9.
\end{equation}
The best-fit values of the parameters are
\cite{SK}
\begin{equation}\label{SKbest}
\Delta m^{2}_{23}=2.5\cdot 10^{-3}\rm{eV}^{2},\quad
\sin^{2}2
\theta_{23}=1.
\end{equation}
The results of the S-K atmospheric neutrino experiment have
been
confirmed by the  K2K \cite{K2K} and MINOS \cite{Minos}
accelerator long-baseline experiments. From the analysis
of the MINOS data was obtained \cite{Minos}:
\begin{equation}\label{Minosdata}
\Delta m^{2}_{23}=(2.38^{+0.20}_{-0.16})\cdot
10^{-3}\rm{eV}^{2},\quad \sin^{2}2 \theta_{23}>0.84~(90\%
~CL).
\end{equation}
From the global analysis of the data of the
 KamLAND reactor experiment and the data of the solar
 neutrino experiments
the following values of the parameters
$\Delta
m^{2}_{12}$ and $\tan^{2} \theta_{12}$ were found
\cite{Kamland}:
\begin{equation}\label{KLsolar}
\Delta m^{2}_{12} = (7.59^{+0.21}_{-0.21})\cdot
10^{-5}~\rm{eV}^{2},\quad\tan^{2} \theta_{12}=
0.47^{+0.06}_{-0.05}~.
\end{equation}

\section{Neutrino production and detection.\\ 
Schr\"odinger
equation for neutrino states}

Let us consider the decay
\begin{equation}\label{decay}
a\to b + l^{+}+\nu_{i}~.
\end{equation}
For the state of the final particles we have
\begin{equation}\label{decay1}
 |f\rangle=\sum_{i}|\nu_{i}\rangle~|l^{+}\rangle
 ~|b\rangle~\langle\nu_{i}l^{+} b~ |S|~a\rangle ~.
\end{equation}
Here $|\nu_{i}\rangle=c_{-1}^{\dag}(p_{i})~|0\rangle$ is the
state of a neutrino with four-momentum $p^{\alpha}_{i}$ and
helicity equal to -1 and $\langle\nu_{i}l^{+} b~
|S|~a\rangle$ is the matrix element of the process
(\ref{decay}).
We are interested in neutrino energies $\gtrsim 1$ MeV. At
such energies, the neutrino mass-squared differences can
safely be neglected in the matrix elements
$\langle\nu_{i}l^{+} b |S|a\rangle$ since $\frac{\Delta
m_{ik}^{2}}{E^{2}}\lesssim 10^{-16} $. Thus, we have
\cite{BilGiunti}
\begin{equation}\label{decay2}
\langle\nu_{i}l^{+}b ~|S|~a\rangle\simeq
U^{*}_{li}~\langle\nu_{l}l^{+}b ~|S|~a\rangle_{\mathrm{SM}}
\end{equation}
Here $\langle\nu_{l}l^{+}b~ |S|~a\rangle_{\mathrm{SM}}$
is the SM matrix element of the production of the  flavor
neutrino
$\nu_{l}$ in the decay
\begin{equation}\label{decay3}
a\to b + l^{+}+\nu_{l}~.
\end{equation}
From (\ref{decay1}) and (\ref{decay2}) we obtain
\begin{equation}\label{decay4}
 |f\rangle\simeq |\nu_{l}\rangle~|l^{+}\rangle~
 |b\rangle~\langle\nu_{l}l^{+} b~
 |S|~a\rangle_{\mathrm{SM}}~,
\end{equation}
where
\begin{equation}\label{decay5}
|\nu_{l}\rangle=\sum^{3}_{i=1}U^{*}_{li}~|\nu_{i}\rangle
\end{equation}
is the state of a flavor neutrino $\nu_{l}$ which is produced
together with the lepton $l^{+}$ in CC weak decays.

We now will consider the propagation of the flavor neutrino
states. In quantum field theory, the evolution equation
is the Schr\"odinger equation
 \begin{equation}\label{schrod}
 \frac{d}{dt}|\Psi(t)\rangle =H ~|\Psi(t)\rangle~,
 \end{equation}
 where $H$ is the total Hamiltonian.

Let us assume that at $t=0$ the flavor neutrino $\nu_{l}$
was produced. Thus, we have $
|\Psi(0)\rangle=|\nu_{l}\rangle$. For the neutrino state in
vacuum at time $t$, we obtain the following expression
\begin{equation}\label{schrod1}
 |\Psi(t)\rangle=\sum_{i}|\nu_{i}\rangle~
 e^{-iE_{i}t}U^{*}_{li},\quad
 E_{i}=\sqrt{p^{2}_{i}+m^{2}_{i}}~.
\end{equation}
If we develop the state $|\Psi(t)\rangle$ on the flavor
states $|\nu_{l'}\rangle$ we have
\begin{equation}\label{schrod2}
    |\Psi(t)\rangle=\sum_{l'}
|\nu_{l'}\rangle~(\sum_{i}U_{l'i}~  e^{-iE_{i}t}~
U^{*}_{li})~.
\end{equation}
Neutrinos are detected via the observation of weak processes.
Let us consider the process
\begin{equation}\label{deepin}
\nu_{l'}+N\to l' +X~.
\end{equation}
Neglecting extremely small terms of the order $\frac{\Delta
m_{ik}^{2}}{E^{2}}$  for the matrix element of the process
$\nu_{i}+N\to l' +X$ we obtain the following relation
\begin{equation}\label{deepin1}
\langle l'X ~ |S|~ \nu_{i}N \rangle\simeq \langle l'X ~  |S|~
\nu_{l'}N \rangle_{\mathrm{SM}}~U_{l'i}~,
\end{equation}
where $\langle l'X  ~ |S|~ \nu_{l'}N \rangle_{\mathrm{SM}}$
is the SM matrix element of the process (\ref{deepin}).
Taking into account the unitarity of the mixing matrix we
find
\begin{equation}\label{deepin2}
\langle l'X ~  |S|~ \nu_{l'}N \rangle=\sum_{i}\langle l'X ~
|S|~ \nu_{i}N \rangle ~ U^{*}_{l'i}\simeq \langle l'X  ~
|S|~ \nu_{l'}N \rangle_{\mathrm{SM}}~.
\end{equation}
Thus, the normalized
$\nu_{l}\to\nu_{l'}$ transition probability is given by the
relation
\begin{equation}\label{transitionprob}
P(\nu_{l}\to\nu_{l'})=
|\sum_{i}U_{l'i}~e^{-i(E_{i}-E_{k})t}~U^{*}_{li}|^{2},
\quad \sum_{l'}P(\nu_{l}\to\nu_{l'})=1~.
\end{equation}
Summarizing, if we neglect terms of the order
$\frac{\Delta m^{2}_{ik}}{E^{2}}\ll 1 $ in the $S$-matrix
elements,  we come to the conclusion that
\begin{itemize}
  \item in weak processes, flavor neutrinos $\nu_{e}$,
      $\nu_{\mu}$ and $\nu_{\tau}$ are produced. The
      states of the flavor neutrinos do not depend on the
      neutrino production (and detection) processes and
      are given by the relation (\ref{decay5}).
\item  the matrix elements of the neutrino production and
    detection processes
are given by the Standard Model.

\item  neutrino oscillations are taking place if the
    neutrino state $ |\Psi(t)\rangle$ is a superposition
    of states of neutrinos with different energies
    (non-stationary states).
\end{itemize}
The non-stationary nature of neutrino oscillations was
advocated by B. Pontecorvo and his collaborators in early
neutrino oscillation papers \cite{Pont,BilPont}.

In the K2K \cite{K2K} and MINOS \cite{Minos} accelerator
experiments, the time of neutrino production and the time of
neutrino detection were measured. In the K2K experiment,
 $\nu_{\mu}$'s were produced in $1.1 ~\mu s $ spills. After
 the time $t\simeq L/c\simeq 0.8\cdot 10^{3}~\mu s$, muon
 neutrinos
 were observed in the Super Kamiokande detector.
Agreement with the S-K results for atmospheric neutrino
oscillations was found.
Thus, it was proven that neutrino oscillations observed in
long-baseline neutrino experiments are a non-stationary
phenomenon.

Here we will address the following questions: Are neutrino
oscillations, in general, a non-stationary phenomenon? Are
stationary neutrino oscillations also possible?

The oscillation phase in equation (\ref{transitionprob}) is
given by the following expression
\begin{equation}\label{phase}
\phi_{ki}=(E_{i}-E_{k})~t\simeq (p_{i}-p_{k})~t+\frac{\Delta
m_{ki}^{2}}{2E}t.    \end{equation}
Generally, we have for the difference of neutrino momenta
\begin{equation}\label{phase1}
(p_{i}-p_{k})\simeq a \frac{\Delta m_{ki}^{2}}{2E}~,
\end{equation}
where  $a\lesssim 1$. Thus, the first term in (\ref{phase})
could be comparable to the second one. We know, however, that
all existing neutrino oscillation data
are described by the expression (\ref{transprob}) in which
the oscillation phase is given by
 \begin{equation}\label{phase2}
\phi_{ki}= \frac{\Delta m_{ki}^{2}}{2E}L
\end{equation}
Comparing (\ref{phase}) and (\ref{phase2}) we come to the
conclusion that in the case of ultrarelativistic neutrinos
with $t\simeq L$ and the Schr\"odinger evolution equation, we
need to assume that neutrinos with different masses in the
mixed flavor state must have the same momenta
($p_{i}=p_{k}=p$). In this case, the flavor neutrino state
$|\nu_{l}\rangle$
is characterized by the momentum $\vec{p}$.

\section{ Wave-function approach to neutrino propagation}
 We now will consider another approach to neutrino
 propagation
 (see, for example Ref. \cite{Giunti} and references
 therein). We will assume that the propagation of  $\nu_{i}$
 with
four-momentum $p^{\alpha}_{i}$  is determined by the Dirac
equation and
that the wave function of the neutrino is the coherent
superposition
\begin{equation}\label{wavefunc}
\Psi_{\nu_{l}}(x)
=\sum_{i}e^{-ip^{\alpha}_{i}x_{\alpha}}U^{*}_{li}~|i\rangle~.
\end{equation}
According to the Dirac equation,
$E_{i}=\sqrt{p^{2}_{i}+m^{2}_{i}}$
and the vector $|i\rangle$ describes a neutrino with mass
$m_{i}$ and helicity equal to -1.

In this case we find the following expression for the
normalized probability of the transition $\nu_{l}\to
\nu_{l'}$
\begin{equation}\label{wavefunc1}
P( \nu_{l}\to \nu_{l'})=|\sum_{i}U_{l'i}~
e^{-i(p^{\alpha}_{i}-p^{\alpha}_{k})x_{\alpha}}~U^{*}_{li}|^{2}.
\end{equation}
It is evident from (\ref{wavefunc1}) that transitions of
flavor neutrinos are due to the fact that wave functions of
neutrinos with different masses at the  distance $\vec{x}$
and after the time $t$ gain  different phases.

Let us consider the oscillation phase. We have
\begin{equation}\label{wavefunc2}
\phi_{ik}=(E_{i}-E_{k})t -(p_{i}-p_{k})x
\end{equation}
For $E_{i}\neq E_{k}$, $p_{i}\neq p_{k}$,
we obtain
\begin{equation}\label{wavefunc3}
\phi_{ik}
\simeq
\frac{\Delta m^{2}_{ik}}{2E}t-(p_{i}-p_{k})(x-t)
\end{equation}
The second term disappears because $x\simeq t$. Thus, if the
neutrino evolution is described by a superposition of plane
waves,
the oscillation phase is given by the standard expression
($\phi_{ik}\simeq
\frac{\Delta m_{ki}^{2}}{2E}L$) in the general case that
$E_{i}\neq E_{k}$, $p_{i}\neq p_{k}$.

Let us now consider the stationary case $E_{i}= E_{k}$. For
the
oscillation phase we obviously obtain the standard
expression
\begin{equation}\label{wavefunc4}
\phi_{ik}= -(p_{i}-p_{k})x=\frac{\Delta m^{2}_{ik}}{2E}L~.
\end{equation}
Thus, if propagating neutrinos are described by a coherent
superposition of plane waves, oscillations are also possible
in the stationary case.

The two principally different possibilities we have discussed
above (evolution of the neutrino state in time or evolution
of the neutrino wave function in space and time) can not be
distinguished by standard neutrino oscillation experiments.
Special neutrino experiments are necessary. Such an
experiment we will discuss in the last part of this paper.

\section{Time-energy uncertainty relation for neutrino
oscillations}

We now discuss the time-energy uncertainty relation for
neutrino os\-cil\-lat\-ions\cite{BilFeilPotz07}.
All uncertainty relations are based on the Cauchy-Schwarz
inequality
\begin{equation}\label{timeenergy}
 \Delta A~\Delta B \geq \frac{1}{2}|\langle
 \Psi|~[A,B]~|\Psi\rangle|~.
\end{equation}
Here $A$ and $B$  are  two Hermitian operators,
$|\Psi\rangle$ is some state
and
\begin{equation}\label{standard}
 \Delta A=
 \sqrt{\overline {A^{2}}- (\bar A )^{2}},\quad \bar
 A=\langle
 \Psi|~A~|\Psi\rangle~.
\end{equation}
The Heisenberg uncertainty relations are  direct consequences
of the commutation relations for  $A$ and $B$. For example,
from the commutation relation $[p,q]=\frac{1}{i}$ follows
that
\begin{equation}\label{uncer}
 \Delta p~\Delta q \geq \frac{1}{2}~.
\end{equation}
Let us stress that the right-hand side of the Heisenberg
uncertainty relations for canonically
conjugated quantities does not depend on the state
$|\Psi\rangle$.

The time-energy uncertainty relation has a completely
different
character. This is connected with the fact that in quantum
theory, time is a parameter and there is no operator of
time.

The time-energy uncertainty relations for different systems
are based on the fact that the
evolution of the quantum systems is determined by the
Hamiltonian. A
general method of deriving the time-energy uncertainty
relations was proposed by Mandelstam and Tamm
\cite{TammMand45}.

Let us consider an operator $O_{H}(t)$ in the Heisenberg
representation.
We have
\begin{equation}\label{timeenergy1}
i~\frac{ d ~O_{H}(t)}{d ~t}=[O_{H}(t),H]~,
\end{equation}
where $H$ is the total Hamiltonian. From the Cauchy-Schwarz
inequality (\ref{timeenergy}) and relation
(\ref{timeenergy1}) we find
\begin{equation}\label{timeenergy2}
\Delta E~\Delta O_{H}(t) \geq \frac{1}{2}
   ~ |\frac{d }{d t} \overline{ O}_{H}(t)|~.
\end{equation}
Here
\begin{equation}\label{average1}
\overline{ O}_{H}(t)= \langle
 \Psi_{H}|~e^{iHt}~O~e^{-iHt}~|\Psi_{H}\rangle=\langle
 \Psi(t)|~O~|\Psi(t)\rangle~,
\end{equation}
where $O$ and $|\Psi(t)\rangle=e^{-iHt}~|\Psi_{H}\rangle$ are
the operator and the vector of the state, respectively, in
the Schr\"odinger representation. It is evident that from
(\ref{timeenergy2}) nontrivial constraints can be obtained
only for non-stationary processes.

In order to obtain the time-energy uncertainty relation for
neutrino oscillations we will choose for  $O$ the operator of
the projection on  the flavor  state $|\nu_{l}\rangle$:
\begin{equation}\label{timeenergy3}
O=|\nu_{l}\rangle~\langle
 \nu_{l}|~.
\end{equation}
Assuming that $|\Psi(0)\rangle=|\nu_{l}\rangle$ we find
\begin{equation}\label{}
\overline{O}_{H}(t)= P_{\nu_{l}\to \nu_{l}}(t)~,
\end{equation}
where $P_{\nu_{l}\to \nu_{l}}(t)$ is the $\nu_{l}$ survival
probability.

If we take into account that $O^{2}=O$, the Mandelstam-Tamm
inequality
(\ref{timeenergy2}) takes the form
\begin{equation}\label{timeenergy4}
\Delta E \geq \frac{1}{2}~\frac{|\frac{d }{d t}P_{\nu_{l}\to
\nu_{l}}(t) |}{\sqrt{P_{\nu_{l}\to
\nu_{l}}(t)-P^{2}_{\nu_{l}\to
\nu_{l}}(t)}}~.
\end{equation}
Let us integrate (\ref{timeenergy4}) over the time $t$ from
$t=0$ to
$t= t_{\rm{1min}}$, where $t_{\rm{1min}}$ is the time at
which the survival probability $P_{\nu_{l}\to \nu_{l}}(t)$
reaches its first minimum.
The time-energy uncertainty relation takes the form
\begin{equation}\label{timeenergy5}
\Delta E~\Delta t \geq
\frac{1}{2}~\left(\frac{\pi}{2}-\arcsin
(2~P_{\nu_{l}\to \nu_{l}}(t_{\rm{1min}})-1)\right),
\end{equation}
where $\Delta t =t_{\rm{1min}}$.

 We will first consider the $\nu_{\mu}\to \nu_{\mu}$
 transition driven by $\Delta m_{23}^{2}$. We have in this
 case
\begin{equation}\label{timeenergy6}
\Delta t_{23}  =t^{(23)}_{\rm{1min}}=2\pi~\frac{E}{\Delta
m_{23}^{2}}~.
\end{equation}
Taking into account that $P_{\nu_{\mu}\to
\nu_{\mu}}(t_{\rm{1min}})\simeq 0$
 we find
\begin{equation}\label{timeenergy7}
\Delta E~\Delta t_{23} \geq \frac{\pi}{2}~.
\end{equation}
From this relation we obtain the following constraint on
$\Delta E$:
\begin{equation}\label{timeenergy8}
\Delta E\geq \frac{\Delta m_{23}^{2}}{4E}~.
\end{equation}
This relation can easily be fulfilled in the atmospheric  and
long-baseline accelerator neutrino experiments.

For the $\bar\nu_{e}\to\bar\nu_{e}$ transition driven by
$\Delta m_{23}^{2}$ we find
\begin{equation}\label{timeenergy9}
 \Delta E\geq \sin2\theta_{13}\frac{\Delta m_{23}^{2}}{2\pi
 E}~.
\end{equation}
Because $\sin^{2}2\theta_{13}\lesssim 2\cdot 10^{-1}$,
relation (\ref{timeenergy9}) gives a much weaker constraint
on $\Delta E$
than (\ref{timeenergy8}).

\section{M\"ossbauer neutrino experiment}
In Refs. \cite{Raghavan1,Raghavan2,WPotzel}, possibilities
have been considered to perform
an experiment on the detection of the tritium electron
antineutrino
with energy $\simeq$ 18.6 keV in the recoilless (M\"ossbauer)
transitions
\begin{equation}\label{recoiless}
^{3}\rm{H}\to ^{3}\rm{He}+\bar\nu_{e},\quad \bar\nu_{e}+
^{3}\rm{He}\to^{3}\rm{H}~.
\end{equation}
In the case of oscillations driven by
$\Delta m_{23}^{2}$ the oscillation length in such an
experiment is equal to
$L^{(23)}_{\mathrm{osc}}\simeq  18.6 ~\mathrm{m}$.

It was estimated in Ref. \cite{Raghavan1} that
\begin{equation}\label{recoiless1}
\Delta E \simeq 8.4\cdot 10^{-12}~\mathrm{eV}
\end{equation}
and the cross section of the resonance absorption of
$\bar\nu_{e}$ in the process $\bar\nu_{e}+
^{3}\rm{He}\to^{3}\rm{H}$ is equal to $\sigma_{R}\simeq
3\cdot 10^{-33}\rm{cm}^{2}$.\footnote {For a more critical
review, see Ref. \cite{Potzel}.}

Let us apply the time-energy uncertainty relation
(\ref{timeenergy9}) to the M\"ossbauer neutrino experiment.
For $\sin^{2}2\theta_{13}=2\cdot 10^{-1}$ (the CHOOZ bound)
we find that the energy uncertainty $\Delta E$ must satisfy
the condition
\begin{equation}\label{timeenergy10}
\Delta E \geq 9\cdot 10^{-9}~\mathrm{eV}~.
\end{equation}
If $\sin^{2}2\theta_{13}= 10^{-2}$ (this value corresponds to
the sensitivity strived for with the future T2K \cite{T2K}
and Daya Bay \cite{Dayabay} experiments) we have
\begin{equation}\label{timeenergy11}
\Delta E \geq 2\cdot 10^{-9}~\mathrm{eV}~.
\end{equation}
Thus, the time-energy uncertainty relation
(\ref{timeenergy9})
is not satisfied in the case of a M\"ossbauer neutrino
experiment
with an
uncertainty of the neutrino energy given by
(\ref{recoiless1}) (see Ref. \cite{Raghavan1}) and even less
so with the much smaller energy uncertainty $\Delta E \simeq
1\cdot 10^{-24}~\mathrm{eV}$ proposed in Ref.
\cite{Raghavan2} (see, however, critical review in Ref.
\cite{Potzel}).
If neutrino oscillations will be observed in
such  an experiment it would mean that the Mandelstam-Tamm
time-energy uncertainty relation is not a characteristic
feature of neutrino oscillations.

\section{Conclusion}
In spite of neutrino oscillations having been observed in
atmospheric, solar, reactor and  long-baseline accelerator
experiments there are several open problems concerning the
theory of neutrino oscillations. In this paper we addressed
the question: is non-stationarity a characteristic feature of
neutrino oscillations?

In reality, this is a question about the propagation of
neutrinos from a source to a detector. We considered here two
fundamentally different options. If the evolution of the
state of flavor neutrinos $|\nu_{l}\rangle$ is determined by
the Schr\"odinger equation for quantum states, neutrino
oscillations are a non-stationary phenomenon. If the
evolution of the emitted neutrinos
$\nu_{i}$ with  momenta $p_{i}$ is determined by the Dirac
equation and the propagating neutrino is described by a
coherent wave function which depends on $\vec{x}$ and $t$,
both non-stationary and stationary neutrino oscillations are
possible.

The standard neutrino-oscillation experiments can not
distinguish between these two possibilities. Special neutrino
experiments are necessary to achieve this. We discussed a
recently proposed tritium/helium-3 M\"ossbauer neutrino
experiment
with practically monochromatic neutrinos.\footnote {For
another discussion of this M\"ossbauer neutrino experiment,
see Ref. \cite{Akhmedov}.}
Such an experiment would be able to decide which assumption
on the neutrino propagation is correct.

Following the Mandelstam-Tamm method we derived the
time-energy uncertainty relation for neutrino oscillations.
In the standard neutrino-oscillation experiments this
relation is satisfied. The M\"ossbauer neutrino experiment
could answer the question whether the Mandelstam-Tamm
time-energy uncertainty relation is
universally applicable to neutrino oscillations.

\section{Acknowledgements}
 It is a pleasure to thank E. Kh. Akhmedov, J. Kopp, and M.
 Lindner for fruitful discussions. S. B. gratefully
 acknowledges the generous hospitality at the
 Physik-Department of the Technische Universit\"at M\"unchen
 in Garching. This work has been supported by funds of the
 Munich Cluster of Excellence (Origin and Structure of the
 Universe), the DFG (Transregio 27: Neutrinos and Beyond),
 and the Maier-Leibnitz-Laboratorium in Garching.

 \end{document}